\documentclass[aps,prd,preprint,nofootinbib]{revtex4}
\usepackage{amsfonts}
\usepackage{mathrsfs}
\usepackage{graphicx}
\usepackage{amsmath}
\usepackage{amssymb}
\usepackage{epsfig}
\usepackage{graphicx}
\usepackage{slashed}
\usepackage{color}
\usepackage{multirow}
\usepackage{ulem}
\usepackage{adjustbox}

\usepackage{stmaryrd}
\allowdisplaybreaks[1]

\graphicspath{{fig/}{dia/}} \DeclareGraphicsExtensions{.eps}
\begin{document}

\title{Isospin violating decays of vector charmonia}

\author {Chao-Qiang Geng$^1$, Chia-Wei Liu$^2$ and Jiabao Zhang$^{1,3}$\footnote{zhangjiabao21@mails.ucas.ac.cn}}
\affiliation{
$^1$School of Fundamental Physics and Mathematical Sciences, Hangzhou Institute for Advanced Study,~UCAS, Hangzhou 310024, China\\
$^2$Tsung-Dao Lee Institute and School of Physics and Astronomy, \\
Shanghai Jiao Tong University, Shanghai 200240, China\\
$^3$Institute of Theoretical Physics,~UCAS, Beijing 100190, China\\
University of Chinese Academy of Sciences, 100190 Beijing, China
}
\date{\today}

\begin{abstract}
We study the isospin violating decays of vector charmonia to $\Lambda\overline{\Sigma}^0$ and its charge conjugate.
They are dominated by the single photon annihilation and can be evaluated reliably with timelike form factors.
We utilize the quark-pair creation model, which is valid for the OZI suppressed decays, to evaluate the form factors.
We obtain the branching fractions of ${\cal B}(J/\psi\to\Lambda\overline{\Sigma}^0+c.c.)=(2.4\pm0.4)\times10^{-5}$ and ${\cal B}(\psi(2\,S)\to\Lambda\overline{\Sigma}^0+c.c.)=(3.0\pm0.5)\times10^{-6}$, which are compatible with the measurements by the BESIII collaborations, respectively.
The decay asymmetries are found to be $\alpha_{J/\psi}=0.314$ and $\alpha_{\psi(2\,S)}=0.461$, which can be examined at  BESIII in the foreseeable future.
\end{abstract}

\maketitle

\section{Introduction}

The decays of vector charmonia $(\psi)$ into baryon and antibaryon have recently been thoroughly studied at BESIII.
On the one hand, the branching fractions and decay asymmetries 
have been precisely measured
~\cite{BESIII:2016nix,BESIII:2020fqg,BESIII:2017kqw,BESIII:2023lkg}.
On the other hand, since the produced baryon-antibaryon pairs are entangled, their sequential decays are utilized as sensitive probes to CP asymmetries~\cite{BESIII:2018cnd,BESIII:2022qax,He:2022jjc} generated by new physics~(NP)~\cite{BESIII:2022rzz,BESIII:2022exh,BESIII:2021ges}.
In Table~\ref{BFalpha}, we list the branching fractions~$({\cal B})$ and decay asymmetries~$(\alpha)$ for $\psi$ decaying to a pair of octet baryon-antibaryons. It is interesting to point out that the measured $\alpha$ between $\Sigma\overline{\Sigma}$ and the others differ in sign, suggesting large breaking effects of the SU(3) flavor symmetry~\cite{Alekseev:2018qjg,Chen:2006yn}, which might attribute to NP.

\begin{table}[htbp]
\caption{The Branching fractions and decay asymmetries of $J/\psi$ and $\psi(2\,S)$ to octet baryon-antibaryons.}
\label{BFalpha}
\begin{tabular}{cccc|cccc}
\hline
\hline
&Channels&$10^{3}{\cal B}$&$\alpha$&&Channels&$10^{4}{\cal B}$&$\alpha$\\
\hline
\multirow{7}{*}{$~~J/\psi~~$}
&$p\bar{p}$&$2.120(29)$&$0.595(19)$&\multirow{7}{*}{$~~\psi(2\,S)~$}&$p\bar{p}$&$2.94(8)$&$1.03(7)$\\
&$n\bar{n}$&$2.09(16)$&$0.50(21)$&&$n\bar{n}$&$3.06(15)$&$0.68(16)$\\
&$\Lambda\overline{\Lambda}$&$1.89(9)$&$0.469(26)$&&$\Lambda\overline{\Lambda}$&$3.81(13)$&$0.824(74)$\\
&$\Sigma^0\overline{\Sigma}^0$&$1.172(32)$&$-0.449(20)$&&$\Sigma^0\overline{\Sigma}^0$&$2.35(9)$&$0.71(11)$\\
&$\Sigma^+\overline{\Sigma}^-$&$1.07(4)$&$-0.5156(68)$&&$\Sigma^+\overline{\Sigma}^-$&$2.43(10)$&$0.682(32)$\\
&$\Xi^0\overline{\Xi}^0$&$1.17(4)$&$0.66(3)$&&$\Xi^0\overline{\Xi}^0$&$2.3(4)$&$0.665(118)$\\
&$\Xi^-\overline{\Xi}^+$&$0.98(4)$&$0.586(16)$&&$\Xi^-\overline{\Xi}^+$&$2.87(11)$&$0.693(69)$\\
\hline
\hline
\end{tabular}
\end{table}

In this work, we focus on the isospin-violating effects.
One way to examine them is to compare the differences among the isospin multiplets. Explicitly, the experimental data of ${\cal B}(J/\psi \to \Xi^- \overline{\Xi}^+ /\Xi^0 \overline{\Xi}^0)$ shows a 10\,\% deviation against the isospin symmetry prediction. 
To further study the isospin violation in baryonic decays, the most direct way is to investigate the decays of vector charmonia $\psi$  to $\Lambda\overline{\Sigma}^0$ and the corresponding charge conjugates, which explicitly violate the isospin symmetry.
In particular, the BESIII collaboration has measured the branching fractions as~\cite{BESIII:2012xdg,BESIII:2021mus}
\begin{equation*}
{\cal B}(J/\psi\to\Lambda\overline{\Sigma}^0+c.c)=(2.83\pm0.23)\times10^{-5},~{\cal B}(\psi(2\,S)\to\Lambda\overline{\Sigma}^0+c.c)=(1.6\pm0.7)\times10^{-6},
\end{equation*}
whereas the CLEO-c collaboration found~\cite{Dobbs:2017hyd}
\begin{equation}
{\cal B}(\psi(2\,S)\to\Lambda\overline{\Sigma}^0+c.c)=(1.23\pm0.24)\times10^{-5},
\end{equation}
which is almost an order of magnitude larger than the BESIII's measurement.
Several theoretical studies have been dedicated to these decay modes~\cite{Claudson:1981fj,Zhu:2015bha,Ferroli:2020xnv,Wei:2009zzh,Jiao:2016syk,Kivel:2022fzk,Mangoni:2022yqq,BaldiniFerroli:2019abd}, where the electromagnetic amplitudes are fitted from the experimental data. 

Generally speaking, the isospin symmetry can be violated by the electric charge and mass difference between $u$ and $d$ quarks. In this work, we only consider the electric charge difference, which is manifested by a single photon exchange amplitude.
As we will see later, 
the amplitudes of $J/\psi\to\gamma^*\to hh'$ with $h^{(\prime)}$ an arbitrary hadron can be calculated by the timelike form factors. 
In this work, we adopt the quark pair creation model~(QPC), also known as the $^{3}P_{0}$ model, to describe the quark-antiquark pair creation from the vacuum~\cite{Micu:1968mk,LeYaouanc:1972vsx,Ackleh:1996yt,Simonov:2011cm}, which may originate from the gluon condensation~\cite{Weber:1988bt}.
The  model has been widely used in the OZI-allowed hadronic decays~\cite{Chen:2007xf,Ke:2011wd,Wang:2013lpa,Gong:2021jkb,Garcia-Tecocoatzi:2022zrf}.
By exploiting it in these OZI-suppressed modes, we provide a direct evaluation of the branching fractions and decay asymmetries of these isospin-violating channels.
All of the decay modes considered in this work can be tested at BESIII.

This paper is organized as follows.
In Sect.~\ref{fomalism}, we show the formalisms which combines the homogeneous bag and quark pair creation models.
The numerical results are given in Sect.~\ref{numerical}.
Section~\ref{conclusion} is the conclusion.

\section{Formalism}\label{fomalism}

\begin{figure}[t!]
\centering
\includegraphics[width=0.45\textwidth]{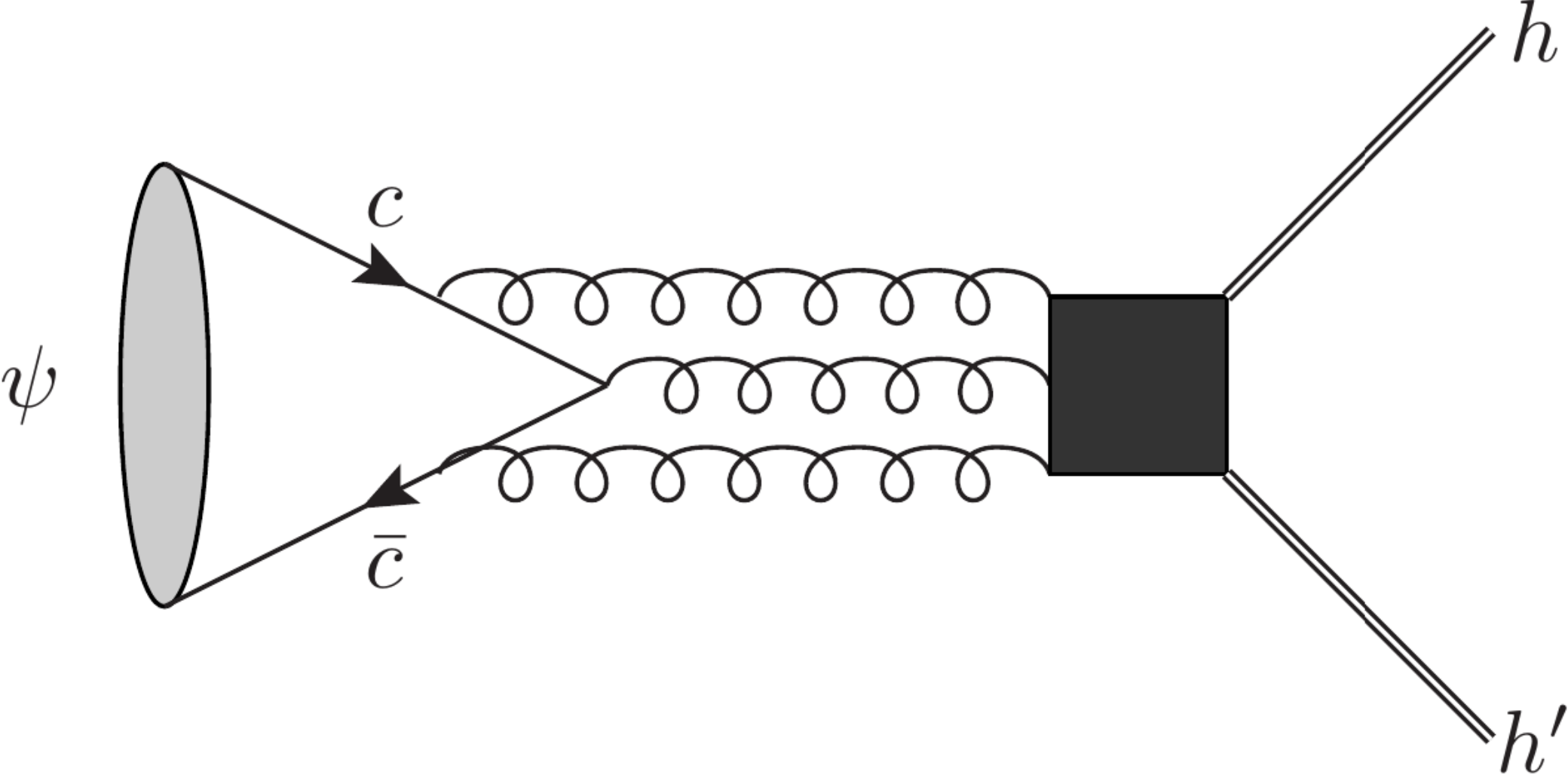}
\includegraphics[width=0.45\textwidth]{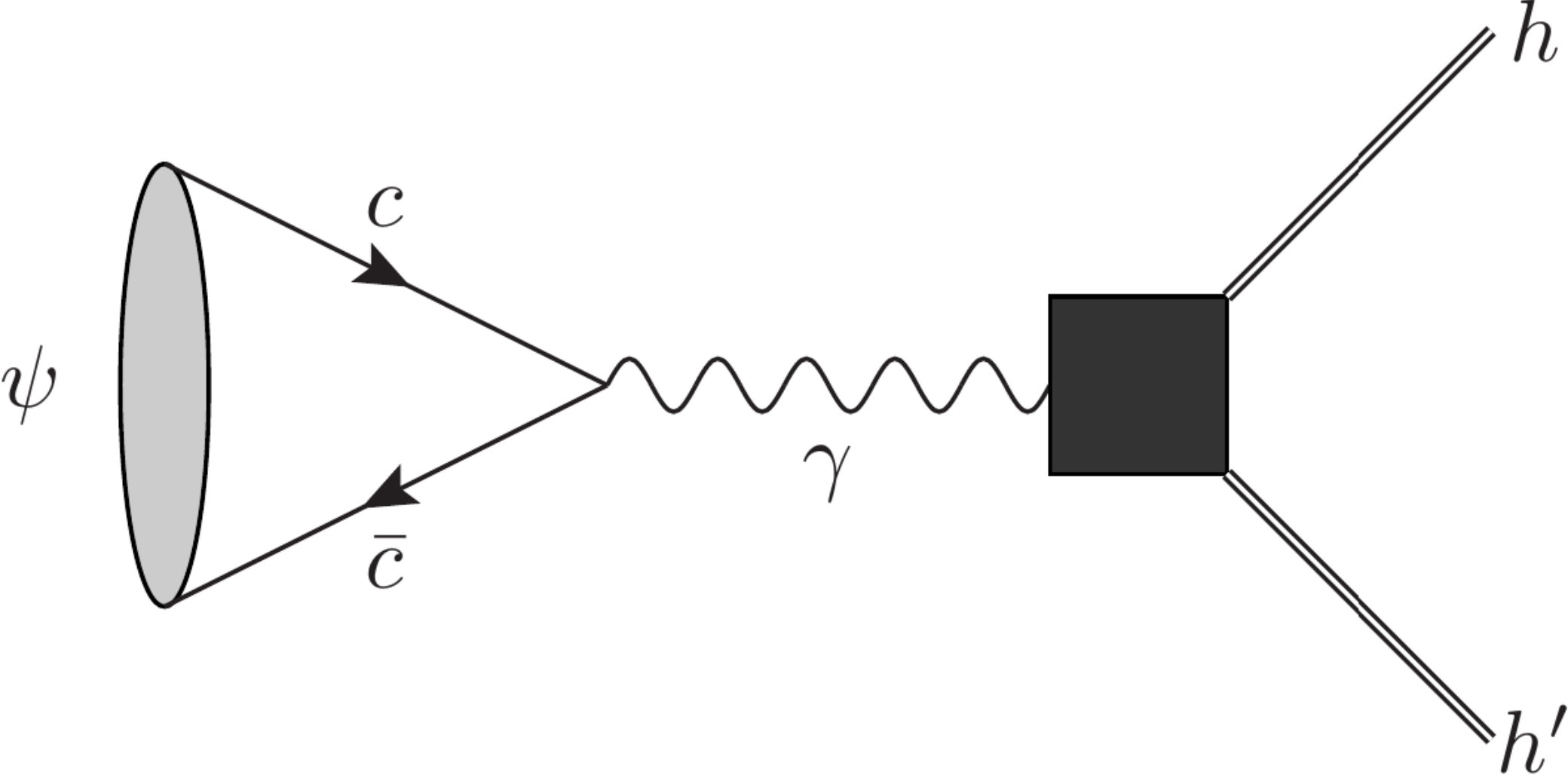}
\caption{The quark diagrams of $\psi\to hh^{\prime}$, where the double lines represent hadrons.}
\label{JpsiFeyn}
\end{figure}

The leading amplitudes of ${\cal A}(\psi\to hh')$ are classified into three categories: ${\cal A}^{ggg}$, ${\cal A}^\gamma$ and ${\cal A}^{gg\gamma}$, where ${\cal A}^{X}$ represents ${\cal A}( \psi \to X \to hh')$.\footnote{${\cal A}^{gg}$ is forbidden by the parity conservation}
In general, the dominant amplitude is ${\cal A}^{ggg}$ since ${\cal A}^{\gamma}/{\cal A}^{ggg}\propto \alpha_{em}/ \alpha_{s}^3\approx 1/2$ with $\alpha_{em(s)}$ being the fine structure constant of QED (QCD).\footnote{
This na\"ive estimation is compatible to the experimental branching fraction of ${\cal B}(J/\psi \to \gamma^* \to \text{hadrons})/{\cal B}(J/\psi \to ggg) = 0.211\pm 0.006$~\cite{ParticleDataGroup:2022pth}.}
Such amplitude is difficult to be evaluated due to the nonperturbative effect of QCD at the charm scale.
In the case of the isospin violating decays, the hierarchy is inverted as ${\cal A}^\gamma \gg {\cal A}^{ggg}$ since the latter is suppressed by the smallness of the mass differenct between $u$ and $d$ quark. It suffices to consider ${\cal A}^\gamma$ solely, as depicted in Fig.~\ref{JpsiFeyn}.

In the following, we will consider the isospin violating decays exclusively and set the quark masses of $m_u$ and $m_d$ to be equivalent. Accordingly, we focus on ${\cal A}^\gamma$ and drop its superscript as confusions are not possible, which ${\cal A}^\gamma$ is decomposed according to the helicities as\footnote{Without lost of generality, we take the velocities of $hh'$ and the polarization of $\psi$ toward $\hat{z}$. For numerical evaluations, we use $\lambda_\psi = \lambda_h -\lambda_{h'}$, where $\lambda_\psi$ is the angular momentum of $\psi$ toward $\hat{z}$. In the Briet frame, the dependence of $\epsilon^\mu(\lambda_\psi)$ on $\lambda_\psi$ is given as $\epsilon_{\mu}(+)=\frac{1}{\sqrt{2}}(0,1,i,0),~\epsilon_{\mu}(-)=\frac{1}{\sqrt{2}}(0,-1,i,0)$ and $\epsilon_{\mu}(0)=(\frac{M_-}{M_\psi}\gamma v,0,0,-\frac{M_+}{M_\psi}\gamma)$, where $M_\pm=M_h\pm M_{h'}$, $\gamma=1/\sqrt{1-v^2}$ and $v$ is the magnitude of the final state velocity.}
\begin{equation}\label{2}
{\cal A}_{\lambda_h\lambda_{h^\prime}}=4\pi\alpha Q_c\frac{f_{\psi}}{M_{\psi}}\sum_{q=u,d,s} Q_qM^q_{\lambda_h\lambda_{h^\prime}},\quad M^{q}_{\lambda_h\lambda_{h^\prime}}=\epsilon_\mu \langle  \lambda_h; \lambda_{h^\prime}|\bar{q}\gamma^\mu q|0\rangle,
\end{equation}
where $Q_q$ is the electric charge of $q$, $\lambda_{h^{(\prime)}}$ is the helicity of $h^{(\prime)}$, and $f_{\psi}$, $M_\psi$ and $\epsilon_{\mu}$ are the decay constant, mass and polarization vector of $\psi$, respectively.
As the isospin has to be violated, the conserved parts of the amplitudes vanish, given by
\begin{equation}
M^u_{\lambda_h\lambda_{h^\prime}}+M^d_{\lambda_h\lambda_{h^\prime}}=M^s_{\lambda_h\lambda_{h^\prime}}=0,
\end{equation}
leading to $\sum_q Q_qM^q_{\lambda_h\lambda_{h^\prime}}=M^u_{\lambda_h\lambda_{h^\prime}}$.
The branching fraction of $\psi\to hh^{\prime}$ is given as
\begin{equation}
{\cal B}(\psi\to hh^\prime)=\frac{1}{3}\frac{|\vec{p}_{h}|}{8\pi M_\psi^2\Gamma_\psi}\sum_{\lambda_{h},\lambda_{h^\prime}}|{\cal A}_{\lambda_{h}\lambda_{h^\prime}}|^2,
\end{equation}
where $\Gamma_\psi$ is the total decay width of $\psi$, $\vec{p}$ is the 3-momentum of $h$ in the rest frame of $\psi$.
For $e^+ e^-\to \psi \to hh' $, there is an additional  parameter in distributions
\begin{gather}
\frac{d\Gamma}{d\cos\theta}\propto1+\alpha\cos^{2}\theta,\quad\alpha=\frac{|{\cal A}_T|^2-2|{\cal A}_L|^2}{|{\cal A}_T|^2+2|{\cal A}_L|^2},
\end{gather}
where $({\cal A}_T, {\cal A}_L)$ correspond to $({\cal A}_{+-},{\cal A}_{++})$ for $\psi \to \Lambda \overline{\Sigma}^0$, and $\theta$ is the angle between the 3-momenta of  $e^+e^-$ and $hh'$. 

In this work, we focus on the baryonic final states $\Lambda\overline{\Sigma}^{0}$ and its charge conjugate.
The matrix element in Eq.~\eqref{2} can be further parametrized by the timelike form factors $(G_E, G_M) $ and $(f_1,f_2)$ as  
\begin{equation}\label{4}
\begin{aligned}
M^{q}_{\lambda_\Lambda\lambda_{\overline{\Sigma}^0}} &=\epsilon_\mu \bar{u} \left[G_M(q^2)\gamma^\mu+\frac{M_+}{q^2}(G_M(q^2)-G_E(q^2))q^\mu\right]v,\\
&=\epsilon_\mu \bar{u} \left[f_1(q^2)\gamma^\mu+f_{2}(q^2)\frac{i\sigma_{\mu\nu}q^{\nu}}{M_+}\right]v,
\end{aligned}
\end{equation}
where $q^\mu=p^\mu+p^{\prime\mu}$ and $p^\mu~(p^{\prime\mu})$ and $u~(v)$ are the 4-momentum and Dirac spinor of $\Lambda~(\overline{\Sigma}^0)$.
The form factors are related to the helicity amplitudes as 
\begin{equation}
{\cal A}_T=\sqrt{2(M_\psi^2-M_-^2)}G_M(s),~{\cal A}_L =\frac{M_+}{M_\psi}\sqrt{M_\psi^2-M_-^2}G_E(s). 
\end{equation}

In this work, we adopt the homogeneous bag model~(HBM) and $^{3}P_{0}$ model to evaluate the electromagnetic baryonic form factors in the timelike region.
Such form factors can be measured with high precision at the BESIII experiment~\cite{Xia:2021agf}, which could provide valuable information for theoretical study.

\subsection{Homogeneous bag model}

In the homogeneous bag model~(HBM), a baryon state is constructed by acting quark field operators on the vacuum state, 
which effectively couples a baryon to quarks by wave functions.  
We take $\Lambda$ baryon as an example, given as 
\begin{equation}\label{eq8}
|\Lambda,\vec{p}=0,\uparrow\rangle=\int[d^3\vec{x}]\frac{1}{\sqrt{6}}\epsilon^{\alpha\beta\gamma}u^\dagger_{a\alpha}(\vec{x}_u)d^\dagger_{b\beta}(\vec{x}_d)s^\dagger_{c\gamma}(\vec{x}_s)\Psi^{abc}_{\uparrow[ud]s}([\vec{x}])|0\rangle,
\end{equation}
where $q^{\dagger}_{a\alpha}(\vec{x})$ is the field operator which creates quark $q$ at position $\vec{x}$, $a$ and $\alpha$ are the spinor and color indices, respectively.
The three quarks are combined into a baryon by the color and spin-flavor-spatial wave function $\epsilon^{\alpha\beta\gamma}$ and $\Psi^{abc}_{[ud]s}$.
We have used the shorthand notation $[\vec{x}]=(\vec{x}_{u},\vec{x}_{d},\vec{x}_{s})$ and $[d^3\vec{x}]=d^3\vec{x}_{u}d^3\vec{x}_{d}d^3\vec{x}_{s}$.
The wave function $\Psi^{abc}_{\uparrow[ud]s}$ for $\Lambda$ baryon is given as
\begin{equation}\label{3}
\Psi^{abc}_{\uparrow[ud]s}([\vec{x}])=\frac{{\cal N}_\Lambda}{\sqrt{2}}\int d^3\vec{x}\left( \phi_{u\uparrow}^a(\vec{x}_u^{\,\prime})\phi_{d\downarrow}^b(\vec{x}_d^{\,\prime})-\phi_{u\downarrow}^a(\vec{x}_u^{\,\prime})\phi_{d\uparrow}^b(\vec{x}_d^{\,\prime})\right) \phi_{s\uparrow}^c(\vec{x}_s^{\,\prime}),
\end{equation}
where $\phi$ is the static bag wave function described in Appendix A,
the subscript $[ud]$ indicates the wave function is antisymmetric in swapping $u$ and $d$ quarks, and 
$\vec{x}'_q = \vec{x}_q - \vec{x}$ is the position of quark $q$ with respect to the bag center $\vec{x}$.

In the original bag model, the hadron state is described by a single bag with its center located at $\vec{x}=0$. However, this configuration is not invariant under Poincar\'e transformations and thus cannot be considered as an eigenstate of four-momentum.
To reconcile the inconsistency, the homogeneous bag model is introduced by duplicating the bag and distributing homogeneously over 
the three-dimensional position space ($\vec{x}$)~\cite{Liu:2022pdk}. 
By construction, 3-dimensional space points are treated equally in Eq.~\eqref{eq8}. 
Such a hadron state is more suitable for describing the decays of hadrons, and has been extensively used in various baryon decays~\cite{Geng:2020ofy}.

As shown in Eq.~\eqref{4}, the decay of $\psi\to\Lambda\overline{\Sigma}^{0}$ are described by the electromagnetic form factors of $\Lambda\overline{\Sigma}^{0}$.
However, the form factors cannot be evaluated even if the hadron wave functions are known. Besides the quark-antiquark pair produced by the photon, two additional pairs of quark-antiquarks are needed in order to form a baryon and an antibaryon.
A possible way to calculate the creation matrix element of the baryon-antibaryon pair is to adopt the crossing symmetry on the hadron level~\cite{Jin:2021onb}.
Nevertheless, it is done by assuming the absence of a singularity in form factors.
In this paper, we adopt the $^{3}P_{0}$ model to describe the creation of quark-antiquark pairs.
By inserting the $^{3}P_{0}$ transition operator, the timelike form factors are directly evaluated.
Details about this model and our approach can be found in the rest of this section.

\subsection{$^{3}P_{0}$ model}

In the $^3P_0$ model, the quark-antiquark pairs are created by the transition operator~\cite{Segovia:2012cd}
\begin{equation}\label{5}
T_q=\sqrt{3}\,\gamma_q \int d^3\vec{x}~:\overline{q}(\vec{x})\,q(\vec{x}):~,
\end{equation}
where $\gamma_q$ is a dimensionless parameter that describes the strength of the creation, and $\sqrt{3}$ is a color factor.
 The $^3P_0$ operator is the simplest effective operator that creates the quark-antiquark pair, which may  originate from the fundamental quantum chromodynamics~(QCD) interaction between quarks and gluons.
The gluon-quark couplings in QCD as well as the condensations are all effectively absorbed into $\gamma_{q}$~\cite{Weber:1988bt}.
Therefore, it is reasonable to expect a universal, model-independent strength parameter $\gamma_{q}$ running with the energy scale as
\begin{equation}
\gamma_{q}(\mu)=\frac{\gamma_{q0}}{\log(\mu/\mu_{0})}.
\end{equation}
In the phenomenological practice, it suffices to fit the parameters $\gamma_{q0}$ and $\mu_{0}$ from various decay experiments, which are 
adopted from Ref.~\cite{Segovia:2012cd} in this work. We emphasize that $\gamma_{q}$ shall not depend on hadron wave functions as it essentially describes the creations of (anti)quarks at the quark level.

In the previous literature, the $^3P_0$ model is  mostly used in the cooperation with 
the  nonrelativistic~(NR) hadron wave functions. At the first glance, it may seem that it conflicts with the bag model, which is essentially a relativistic quark model. However,
 the relativistic corrections in $\psi\to\Lambda\overline{\Sigma}^{0}$ are rather small and the HBM has a well-defined NR limit. We also present the results in the NR limit in Sect.~\ref{numerical}.

With these phenomenological models, the matrix element is now given as
\begin{equation}\label{eq8}
\begin{aligned}
M^{u}_{\lambda_\Lambda\lambda_{\overline{\Sigma}^0}}
&=\epsilon_\mu \langle \lambda_\Lambda; \lambda_{\overline{\Sigma}^0}|\bar{u}\gamma^\mu u T_dT_s |0\rangle\\
&=3\gamma_q^2 
\sum_{[\lambda]} N^{\lambda_\Lambda\lambda_{\overline{\Sigma}^0}} ([\lambda]) \int d^3\vec{x}_\Delta \Gamma^{\lambda_ \psi}_{\lambda_{u}\lambda_{\bar{u}}}(\vec{x}_\Delta)
E_{\lambda_{d}\lambda_{\bar{d}}}(\vec{x}_\Delta)E_{\lambda_{s}\lambda_{\bar{s}}}(\vec{x}_\Delta),
\end{aligned}
\end{equation}
where $[\lambda]$ collects all the quark spins and 
$N^{\lambda_\Lambda\lambda_{\overline{\Sigma}^0}}([\lambda])$ is the spin-flavor overlapping.
The integration over $\vec{x}_{\Delta}$ is directly related to the integration over all the bag centers $\vec{x}$ in Eq.~\eqref{3}, which distinguishes the HBM with the original bag model.
The vertex functions of  $\Gamma^{\psi}_{\lambda_u\lambda_{\bar{u}}}$ and  $E_{\lambda_q\lambda_{\bar{q}}}$ correspond to the productions of the quark-antiquark pairs due to the QED vertex and $T_q$, respectively, given as
\begin{equation}\label{vf1}
\begin{aligned}
&\Gamma^{\lambda_ \psi}_{\lambda_{u}\lambda_{\bar{u}}}(\vec{x}_\Delta)
= \int d^3 \vec{x}_u {\cal G}
^{\lambda_ \psi}_{\lambda_{u}\lambda_{\bar{u}}}
=\int d^3\vec{x}_{u}\phi^\dagger_{u\lambda_{u}}(\vec{x}_{u}+\frac{1}{2}\vec{x}_\Delta)\Upsilon \tilde{\phi}^*_{\bar{u}\lambda_{\bar{u}}}(\vec{x}_{u}-\frac{1}{2}\vec{x}_\Delta),\\
&E_{\lambda_{q}\lambda_{\bar{q}}}(\vec{x}_\Delta)= \int d^3 \vec{x}_q {\cal E}
_{\lambda_{q}\lambda_{\bar{q}}}
=
\frac{1}{\gamma}\int d^3\vec{x}_{q}\phi^\dagger_{q\lambda_{q}}(\vec{x}_{q}+\frac{1}{2}\vec{x}_\Delta){\cal S}\tilde{\phi}^*_{\bar{q}\lambda_{\bar{q}}}(\vec{x}_{q}-\frac{1}{2}\vec{x}_\Delta),
\end{aligned}
\end{equation}
where $\Upsilon= S_v\gamma^0\epsilon_\mu \gamma^\mu S_{-v}$, ${\cal S}\equiv S_v\gamma^0S_{-v}$ and $S_{\pm v}=(\sqrt{\gamma+1}\pm\sqrt{\gamma-1}\gamma^{0}\gamma^{3})/\sqrt{2}$ boost the wave function towards $\pm z$ direction,\footnote{
At the risk of abuse of notation, we adopt the conventional definition of $\gamma = 1/ \sqrt{1-v^2}$ in $S_{\pm v}$, which shall not be confused with the couplings of $\gamma_q$ in $T_q$. 
} and 
$\tilde{\phi}$ is the charge conjugation of the wave function $\phi$.
Note that there is no spectator quark, which  clearly differs from the form factors at the spacelike region.
Besides, without introducing $^{3}P_{0}$ operators, we have ${\cal S}=1$ in Eq.~\eqref{vf1}, leading to vanishing $E_{\lambda_{q}\lambda_{\bar{q}}}(\vec{x}_\Delta)$ for arbitrary spin configurations.

As we can see from Eqs.~\eqref{eq8} and \eqref{vf1}, one has to perform a twelve-fold integral to obtain the final result.
After choosing the appropriate coordinates for different integrals, we manage to reduce the complexity of the calculation significantly,   described in  Appendix A.
By plugging Eq.~\eqref{wf}, we find that 
\begin{eqnarray}\label{Bmtrx}
&&{\cal A}_{T}={\cal C}_B \langle 2\Gamma^+_{--}E_{+-}^{d}E_{++}^{s}-\Gamma^+_{-+}E_{+-}^{d}E_{+-}^{s}+\Gamma^+_{+-}E_{-+}^{d}E_{+-}^{s}-2\Gamma^+_{+-}E_{--}^{d}E_{++}^{s}\rangle,\nonumber\\
&&{\cal A}_{L}={\cal C}_B\langle-2\Gamma^0_{-+}E_{++}^{d}E_{+-}^{s}+2\Gamma^0_{++}E_{-+}^{d}E_{+-}^{s}\rangle,\nonumber\\
&&{\cal C}_B=4\pi\alpha Q_{c}\frac{f_{\psi}}{M_{\psi}}3\gamma_q^2{\cal N}_\Lambda{\cal N}_{\overline{\Sigma}^0}\frac{1}{2\sqrt{3}}\,,
\end{eqnarray}
where ${\cal N}$ is the normalization constant and $\langle \cdots \rangle$ stand for $\int d^3 \vec{x}_\Delta$.

\section{Numerical results}\label{numerical}

We extract $f_\psi$ through the experiments of 
${\cal B}( \psi\to e^+e^-) $ and find that  $f_{J/\psi}=416$ MeV and $f_{\psi(2\,S)}=294$ MeV.
The bag radius of 
$\Lambda$ and $\overline{\Sigma}^0$ are taken to be $5\,\text{GeV}^{-1}$.
The running of $\gamma_q(\mu) $ is taken from Ref.~\cite{Segovia:2012cd}, fitted from the decay widths of heavy mesons, where $\mu$ is the energy scale.
To be conservative, we consider $10\%$ variations of $\mu$, leading to $\gamma_q=0.295(14)$ for $J/\psi$ and $0.278(13)$ for $\psi(2\,S)$.

\begin{table}[htbp]
\caption{The form factors for $J/\psi,\psi(2\,S)\to\Lambda\overline{\Sigma}^{0^{}}$.}
\label{formfactors}
\begin{tabular}{cccccc}
\hline
\hline
&$~G_E(M_\psi^2)~$&$~G_M(M_\psi^2)~$&$~f_1(M_\psi^2)~$&$~f_2(M_\psi^2)~$\\
\hline
$J/\psi$&$~0.0151(14)~$&$~-0.0155(15)~$&$~-0.0538(51)~$&$~0.0383(36)~$\\
$\psi(2\,S)$&$~0.0132(12)~$&$~-0.0239(22)~$&$~-0.0479(45)~$&$~0.0240(22)~$\\
\hline
\hline
\end{tabular}
\end{table}

\begin{figure}[htbp]
\centering
\includegraphics[width=0.48\textwidth]{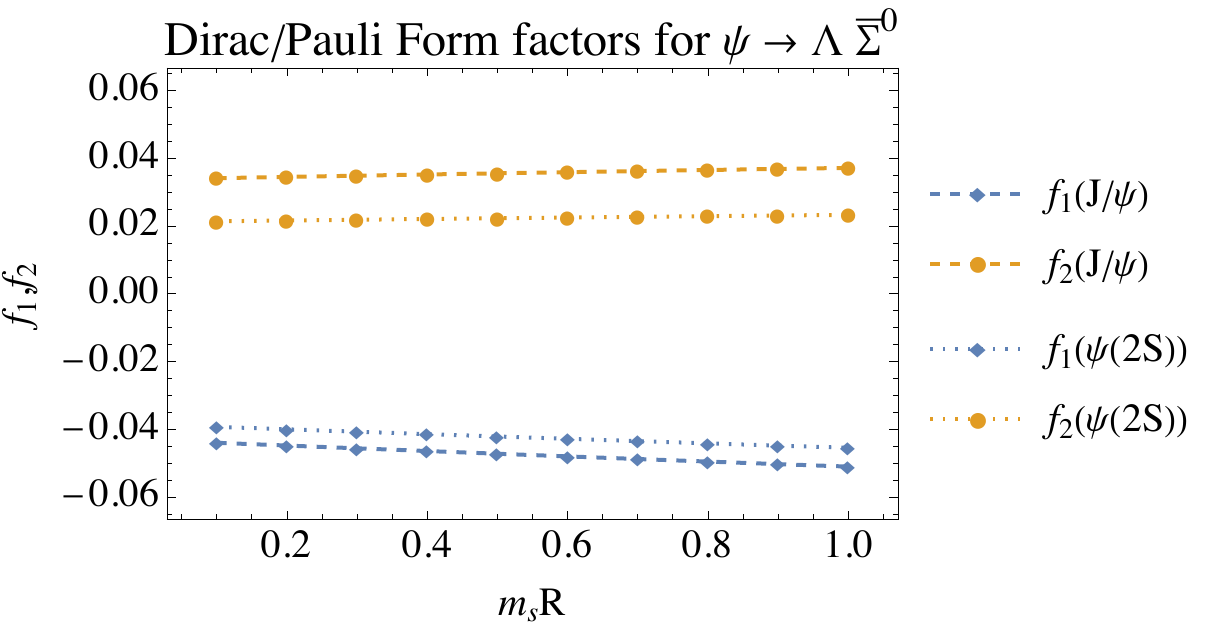}
\includegraphics[width=0.48\textwidth]{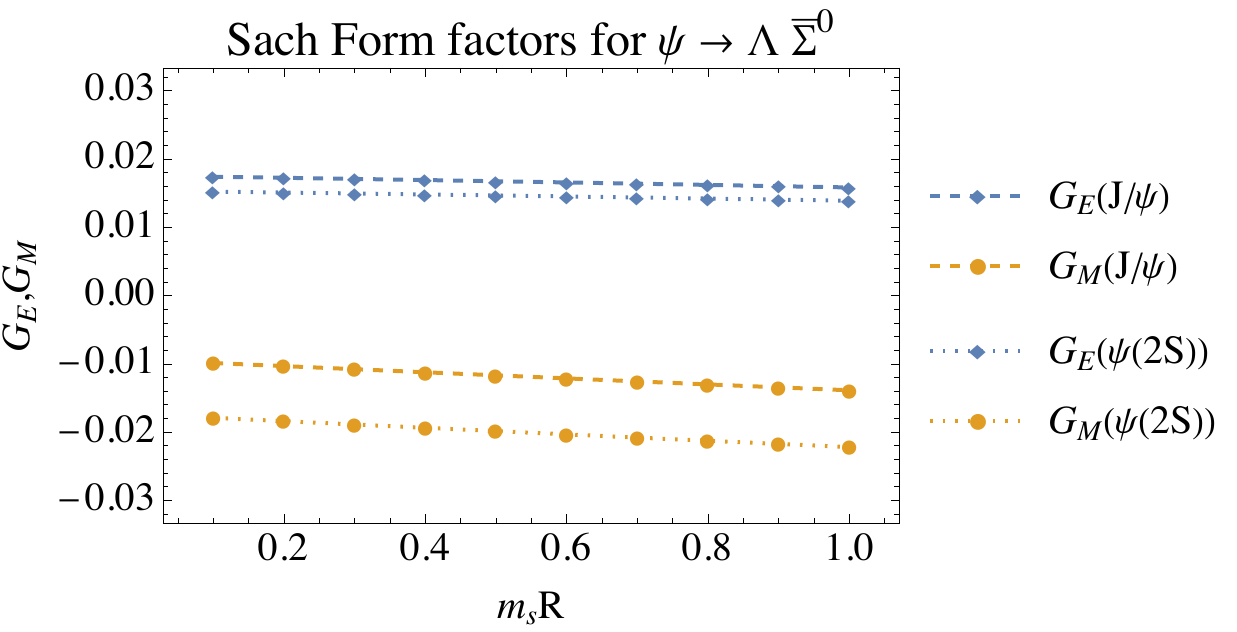}
\caption{The form factors of $\psi\to\Lambda\overline{\Sigma}^0$ versus the dimensionless parameter $m_sR$ from the HBM.}
\label{FFRms}
\end{figure}

Remarkably, the numerical results depend little on the bag radius.
The numerical results of the timelike form factors $(G_E, G_M) $ and $(f_1,f_2)$ are listed in Table~\ref{formfactors}.
There are only two dimensionful parameters in the model, which correspond to the bag radius $R$ and the strange quark mass $m_{s}$. 
The form factors are dimensionless and thus depend only on $m_sR$, which vanishes in the $SU(3)_F$ limit. 
We plot the form factors versus $m_sR $
in Fig.~\ref{FFRms}, which shows slight dependency. 

In Table~\ref{below}, we present the branching fractions and decay asymmetries.
The dependence on $\gamma_q$ is canceled in $\alpha_{\psi}$, leading to negligible uncertainties on $\alpha_\psi$.

\begin{table}[htbp]
\caption{The branching fractions and decay asymmetries of $\psi\to hh^{\prime}$.}
\label{below}
\begin{tabular}{ccccc}
\hline
\hline
Channels&\multicolumn{2}{c}{$10^{5}\cal B$}&$\alpha_{\psi}$\\
\hline
&Exp.&This work&This work\\
\hline
$~~J/\psi\to\Lambda\overline{\Sigma}^0+c.c.~~$&$~~2.83\pm0.23~~$~\cite{BESIII:2012xdg}&$~~2.40\pm0.40~~$&$0.314$\\
\hline
\multirow{2}{*}{$~~\psi(2\,S)\to\Lambda\overline{\Sigma}^0+c.c.~~$}&$~~0.16\pm0.07~~$~\cite{BESIII:2021mus}&$~~0.30\pm0.05~~$&$0.461$\\
&$~~1.23\pm0.24~~$~\cite{Dobbs:2017hyd}&$~~0.75\pm0.13~~$&$0.786$\\
\hline
\hline
\end{tabular}
\end{table}

The predicted ${\cal B}$ of $J/\psi\to\Lambda\overline{\Sigma}^0+c.c.$ is consistent with the BESIII measurements, whereas the ones of $\psi(2\,S)$ sits between the  experimental measurements at the BESIII~\cite{BESIII:2021mus} and CLEO~\cite{Dobbs:2017hyd} collaborations.
From Table~\ref{formfactors}, we can see that $|G_M|$ for $\psi(2\,S)$ is larger than that for $J/\psi$, which contradicts the common belief that form factors should decrease as $q^2$ increases.
In this work, we consider two scenarios.
In Table~\ref{below},
the upper row results of $\psi(2\,S)$ 
are evaluated by taking 
$G_{E,M}( M_{\psi(2\,S)}^2 ) = G_{E,M}( M_{J/\psi }^2 )$, while the lower row results by calculating directly within the $^3P_0$ model.
The first scenario favors ${\cal B}$ measured at BESIII, while the second at CLEO-c. 
We note that the branching ratio between $J/\psi$ and $\psi(2\,S)$ in the  first scenario is compatible with the na\"ive expectation of $ {\cal B}_{\psi(2\,S)}^{ee}/{\cal B}_{J/\psi}^{ee}\approx 13\% $ with ${\cal B}_\psi^{ee}$ the branching fraction of $\psi\to e^+e^-$.

To examine the results, we consider the nonrelativistic~(NR) limit by taking $m_{q} \to \infty$.
As a result, the terms in Eq.~\eqref{Bmtrx} with $E_{\pm\mp}$ vanish in the NR limit, resulting in ${\cal A}_{T}={\cal C}_B \langle-2\Gamma^+_{+-}E_{--}^{d}E_{++}^{s}\rangle$, ${\cal A}_L=0$. Note that this is indeed the interpretation of the $^{3}P_{0}$ quantum number, in which the created $q\bar{q}$ has the spin configuration of $|\uparrow\downarrow+\downarrow\uparrow\rangle$.
When the relativistic corrections are included, other spin configurations also contribute, leading to the amplitude shown in Eq.~\eqref{Bmtrx}.

\section{Conclusions}\label{conclusion}

In this work, we study the isospin violating decays of vector charmonia $\psi$ in both baryonic and mesonic sectors.
Such decays  are attributed to the single photon annihilation and suppressed by the OZI rule.
We utilize the $^{3}P_{0}$ model to calculate the timelike form factors.

The branching fractions of $\psi\to\Lambda\overline{\Sigma}^0+c.c.$ are prediceted  as $2.4(4)\times10^{-5}$ for $J/\psi$ and $0.30(5)\times10^{-5}$ for $\psi(2\,S)$, which are both consistent with the experimental measurements at BESIII.
For the decay asymmetries, we predict $\alpha_{J/\psi}=0.314$ and $\alpha_{\psi(2\,S)}=0.461$ for $\psi\to \Lambda \overline{\Sigma}^0$, which can be tested at BESIII in the foreseeable future.

\section*{Acknowledgments}
We would like to express our sincere appreciation to Prof. Xiaorong Zhou and Zekun Jia for their valuable insights during the development of this work.
This work is supported in part by the National Key Research and Development Program of China under Grant No. 2020YFC2201501 and  the National Natural Science Foundation of China (NSFC) under Grant No. 12147103 and 12205063.

\appendix 
\section{Wave functions and bag integrals in HBM}

The quark  and antiquark bag wave functions are given as
\begin{equation}\label{wf}
\phi_{q}=\begin{pmatrix}u\,\chi\\iv\,\hat{r}\cdot\vec{\sigma}\chi\end{pmatrix},
~\tilde{\phi}_{\bar{q}}=i\gamma^2\phi_{q},
\end{equation}
respectively,
where $u=\sqrt{E_q+m_q}j_0(p_qr),v=\sqrt{E_q-m_q}j_1(p_qr)$, $\chi$ is the usual Pauli spinor, with $E_q=\sqrt{m_q^2+p_q^2}$, $\chi_\uparrow=(1,0)^T$ and $\chi_\downarrow=(0,1)^T$.
The spatial distributions are governed by
 the zeroth and first spherical Bessel functions of $j_{0,1}(p_qr)$, where $p_{q}$ is the quantized 3-momentum
\begin{equation}
\tan(p_q R)=\frac{p_q R}{1-m_q R-E_q R}\,,
\end{equation}
and  $R$ is the bag radius of the hadron, fitted from the mass spectrum.

We adopt the following normalization condition for the baryon states
\begin{equation}
\langle\vec{\Lambda}, \lambda_\Lambda \mid \vec{\Lambda}^{\prime}, \lambda_\Lambda^{\prime}\rangle=u_\Lambda^{\dagger} u_\Lambda^{\prime}(2 \pi)^3 \delta^3\left(\vec{p}_{\Lambda}-\vec{p}\,^{\prime}_{\Lambda}\right),
\end{equation}
where $u_{\Lambda},\lambda_{\Lambda}$ are the spinor and the spin of the $\Lambda$ baryon, its normalization factor is found to be
\begin{equation}
{\cal N}_\Lambda=\left(\frac{1}{\bar{u}_\Lambda u_\Lambda} \int d^3 \vec{x}_{\Delta} \prod_{q=u,d,s} D^{q}\left(\vec{x}_{\Delta}\right)\right)^{-1/2},
\end{equation}
with
\begin{equation}\label{A6}
D^{q}(\vec{x}_\Delta)=\int d^{3}\,\vec{x}_{q}{\cal D}^{q}=\int d^3\vec{x}_{q}\phi^\dagger_{q}\left(\vec{x}_{q}+\frac{1}{2}\vec{x}_\Delta\right){\phi}_{q}\left(\vec{x}_{q}-\frac{1}{2}\vec{x}_\Delta\right).
\end{equation}
Note that $D^{q}(\vec{x}_\Delta)$ is independent of the velocity and spin of the baryon.

\begin{figure}
    \centering
    \includegraphics[width=0.75\textwidth]{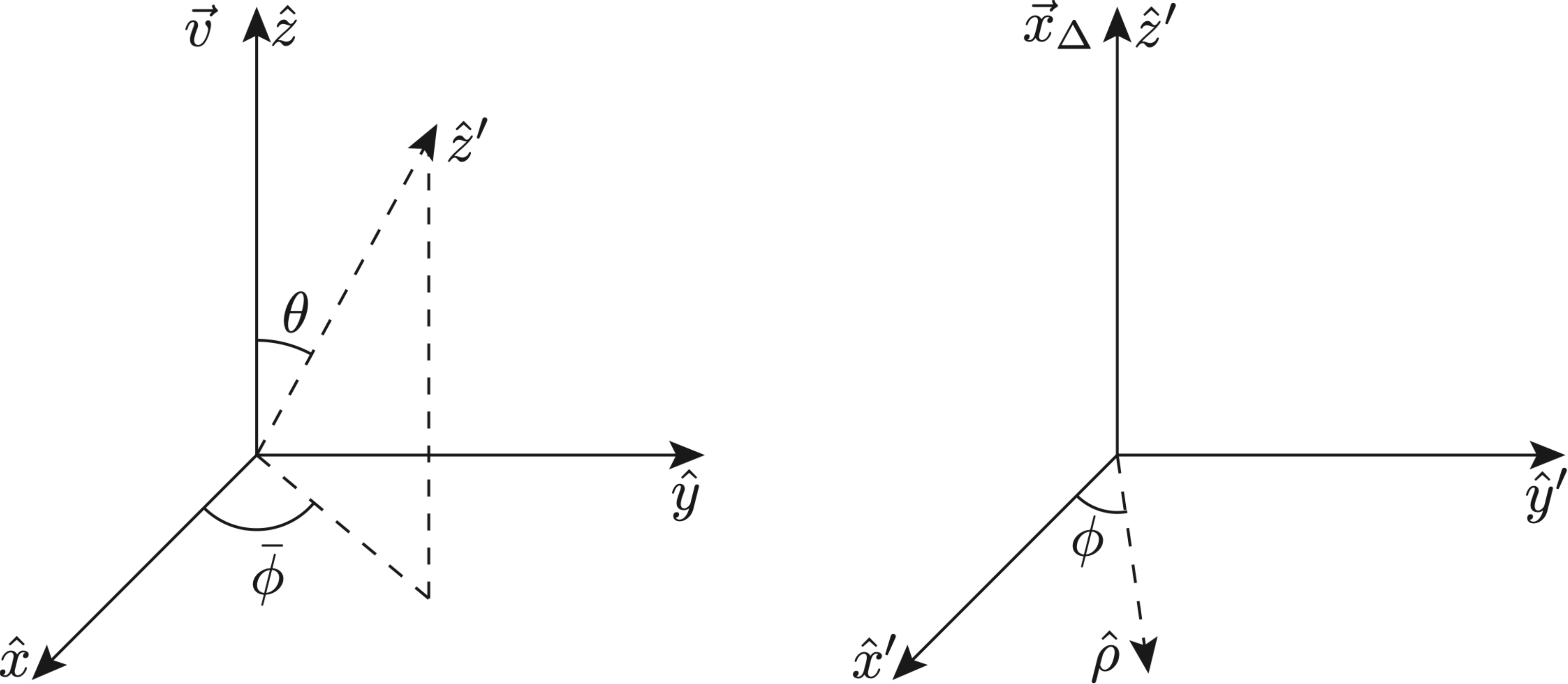}
    \caption{The cylindrical coordinate, where 
$\hat{z}$ and $\hat{z}'$ are chosen to be parallel to  $\vec{v}$ and $\vec{x}_\Delta$, repsectively.}
    \label{cylindrical}
\end{figure}

To evaluate Eqs.~\eqref{vf1} and \eqref{A6},
we use the coordinate shown in Fig.~\ref{cylindrical}, where 
$\hat{z}$ and $\hat{z}'$ are chosen to be parallel to  $\vec{v}$ and $\vec{x}_\Delta$, respectively.
By changing the integration variables from $ d^3\vec{x}_q$ to $d\rho dz' d \phi$ and integrating over $d \phi$, we arrive at 
\begin{equation}
\int {\cal D}^{q}d\phi=2\pi\left({\cal E}_{2}+{\cal E}_{3}+2{\cal E}_{4}\right),
\end{equation}
\begin{equation}
\begin{aligned}
\int{\cal E}_{\pm\mp}d\phi=\mp2\pi e^{\mp i\tilde{\phi}}\sin\theta\left(i{\cal E}_{1}+2v\cos\theta{\cal E}_{3}-2v\cos\theta{\cal E}_{4}\right),\\
\int{\cal E}_{\pm\pm}d\phi=2\pi(i\cos\theta{\cal E}_{1}+v{\cal E}_{2}+v\cos2\theta{\cal E}_{3}-2v\cos^{2}\theta{\cal E}_{4}),
\end{aligned}
\end{equation}
\begin{gather}
\int {\cal G}^{\pm}_{\pm\mp}d\phi
	=2\sqrt{2}\pi\gamma\left[iv\cos\theta{\cal E}_{1}+{\cal E}_{2}+\cos^{2}\theta{\cal E}_{3}+\sin^{2}\theta{\cal E}_{4}\right],\nonumber\\
	\int {\cal G}^{\pm}_{\mp\pm}d\phi
	=2\sqrt{2}\pi\gamma e^{\pm2i\tilde{\phi}}\sin^2\theta({\cal E}_{3}-{\cal E}_{4}),\nonumber\\
    \int {\cal G}^{\pm}_{\pm\pm}d\phi=\int {\cal G}^{\pm}_{\mp\pm}d\phi=\sqrt{2}\pi\gamma e^{\pm i\tilde{\phi}}\sin\theta\left[\pm iv{\cal E}_{1}+2\cos\theta({\cal E}_{3}-{\cal E}_{4})\right],\\
    \int {\cal G}^{3}_{\pm\mp}d\phi
	=\mp2\pi e^{\mp i\tilde{\phi}}\sin2\theta\left({\cal E}_{3}-{\cal E}_{4}\right),\nonumber\\
    \int {\cal G}^{3}_{\pm\pm}d\phi
	=2\pi\left(-{\cal E}_{2}+\cos2\theta{\cal E}_{3}-2\cos^{2}\theta{\cal E}_{4}\right),.\nonumber
\end{gather}
where
\begin{equation}
	\begin{aligned}
		{\cal E}_{1}&=\frac{z_{-}}{r_{-}}u^{+}v^{-}-\frac{z_{+}}{r_{+}}u^{-}v^{+}, & {\cal E}_{2}=u^{+}u^{-},&\\
		{\cal E}_{3}&=z_{+}z_{-}\frac{v^{+}v^{-}}{r_{+}r_{-}}, & {\cal E}_{4}=\frac{1}{2}\rho^{2}\frac{v^{+}v^{-}}{r_{+}r_{-}},&\\
	\end{aligned}
\end{equation}
 $u^\pm=u(\vec{x}\pm\vec{x}_\Delta/2)$, $v^\pm=v(\vec{x}\pm\vec{x}_\Delta/2)$,  $z_\pm=z\pm|\vec{x}_\Delta|/2$ and $r_\pm=\sqrt{z_\pm^2+\rho^2}$.

To obtain $N^{\lambda_\Lambda \lambda_{\overline{\Sigma}^0}}$,
the spin-flavor parts of $\Lambda$ and $\overline{\Sigma}^0$ wave functions are
\begin{gather}
    |\overline{\Sigma}^0,\uparrow\rangle=\frac{1}{\sqrt{6}}\left(-s_{\uparrow}d_{\downarrow}u_{\uparrow}-s_{\uparrow}d_{\uparrow}u_{\downarrow}+2s_{\downarrow}d_{\uparrow}u_{\uparrow}\right)|0\rangle,\nonumber\\
    |\overline{\Sigma}^0,\downarrow\rangle=\frac{1}{\sqrt{6}}\left(s_{\downarrow}d_{\uparrow}u_{\downarrow}+s_{\downarrow}d_{\downarrow}u_{\uparrow}-2s_{\uparrow}d_{\downarrow}u_{\downarrow}\right)|0\rangle,\\
    |\Lambda,\updownarrow\rangle=\frac{1}{\sqrt{2}}\left(
d^{\dagger}_{\downarrow}u^{\dagger}_{\uparrow}-d^{\dagger}_{\uparrow}u^{\dagger}_{\downarrow}\right)s^{\dagger}_{\updownarrow}|0\rangle.\nonumber
\end{gather}
Plugging them into Eq.~\eqref{eq8}, we find
\begin{equation}
\begin{aligned}
{\cal A}_{T}\propto\frac{1}{2\sqrt{3}}&\langle-\Gamma^+_{--}E_{++}^{d}E_{+-}^{s}+2\Gamma^+_{--}E_{+-}^{d}E_{++}^{s}-\Gamma^+_{-+}E_{+-}^{d}E_{+-}^{s}\\
&+\Gamma^+_{+-}E_{-+}^{d}E_{+-}^{s}-2\Gamma^+_{+-}E_{--}^{d}E_{++}^{s}+\Gamma^+_{++}E_{--}^{d}E_{+-}^{s}\rangle,\\
{\cal A}_{L}\propto\frac{1}{2\sqrt{3}}&\langle+\Gamma^0_{-+}E_{+-}^{d}E_{++}^{s}+\Gamma^0_{--}E_{++}^{d}E_{++}^{s}-2\Gamma^0_{-+}E_{++}^{d}E_{+-}^{s}\\
&-\Gamma^0_{++}E_{--}^{d}E_{++}^{s}-\Gamma^0_{+-}E_{-+}^{d}E_{++}^{s}+2\Gamma^0_{++}E_{-+}^{d}E_{+-}^{s}\rangle.
\end{aligned}
\end{equation}
Due to the parity conservation, the amplitudes are invariant under the transformation $(\lambda_\psi,\lambda_q, \lambda_{\overline{q}}) \to (- \lambda_\psi, - \lambda_q, -\lambda_{\overline{q}}) $, leading to
\begin{gather}
\langle-\Gamma^+_{--}E^u_{++}E^s_{+-}+\Gamma^+_{++}E^u_{--}E^s_{+-}\rangle=0,\nonumber\\
\langle-\Gamma^0_{++}E^u_{--}E^s_{++}+\Gamma^0_{--}E^u_{++}E^s_{++}\rangle=0,\\
\langle-\Gamma^0_{+-}E^u_{-+}E^s_{++}+\Gamma^0_{-+}E^u_{+-}E^s_{++}\rangle=0.\nonumber
\end{gather}


\begin{thebibliography}{9}

\bibitem{BESIII:2016nix}
M.~Ablikim \textit{et al.} [BESIII],
Phys. Lett. B \textbf{770}, 217-225 (2017).

\bibitem{BESIII:2020fqg}
M.~Ablikim \textit{et al.} [BESIII],
Phys. Rev. Lett. \textbf{125}, 052004 (2020).

\bibitem{BESIII:2017kqw}
M.~Ablikim \textit{et al.} [BESIII],
Phys. Rev. D \textbf{95}, 052003 (2017);\textbf{106}, L091101 (2022)

\bibitem{BESIII:2023lkg}
M.~Ablikim \textit{et al.} [BESIII],
arXiv:2302.09767;2302.13568;2304.14655.

\bibitem{BESIII:2018cnd}
M.~Ablikim \textit{et al.} [BESIII],
Nature Phys. \textbf{15}, 631-634 (2019).

\bibitem{BESIII:2022qax}
M.~Ablikim \textit{et al.} [BESIII],
Phys. Rev. Lett. \textbf{129}, 131801 (2022).

\bibitem{He:2022jjc}
X.~G.~He and J.~P.~Ma,
Phys. Lett. B \textbf{839}, 137834 (2023).

\bibitem{BESIII:2022rzz}
M.~Ablikim \textit{et al.} [BESIII],
Phys. Lett. B \textbf{838}, 137698 (2023);\textbf{839}, 137785 (2023)

\bibitem{BESIII:2021ges}
M.~Ablikim \textit{et al.} [BESIII],
Phys. Rev. D \textbf{105}, 012008 (2022);\textbf{106}, 072008 (2022)

\bibitem{BESIII:2022exh}
M.~Ablikim \textit{et al.} [BESIII],
Sci. China Phys. Mech. Astron. \textbf{66}, 221011 (2023).

\bibitem{Alekseev:2018qjg}
M.~Alekseev, A.~Amoroso, R.~B.~Ferroli, I.~Balossino, M.~Bertani, D.~Bettoni, F.~Bianchi, J.~Chai, G.~Cibinetto and F.~Cossio, \textit{et al.}
Chin. Phys. C \textbf{43}, 023103 (2019).

\bibitem{Chen:2006yn}
H.~Chen and R.~G.~Ping,
Phys. Lett. B \textbf{644}, 54-58 (2007).

\bibitem{BESIII:2012xdg}
M.~Ablikim \textit{et al.} [BESIII],
Phys. Rev. D \textbf{86}, 032008 (2012).

\bibitem{BESIII:2021mus}
M.~Ablikim \textit{et al.} [BESIII],
Phys. Rev. D \textbf{103}, 112004 (2021).

\bibitem{Dobbs:2017hyd}
S.~Dobbs, K.~K.~Seth, A.~Tomaradze, T.~Xiao and G.~Bonvicini,
Phys. Rev. D \textbf{96}, 092004 (2017).

\bibitem{Claudson:1981fj}
M.~Claudson, S.~L.~Glashow and M.~B.~Wise,
Phys. Rev. D \textbf{25}, 1345 (1982).

\bibitem{Zhu:2015bha}
K.~Zhu, X.~H.~Mo and C.~Z.~Yuan,
Int. J. Mod. Phys. A \textbf{30}, 1550148 (2015).

\bibitem{Ferroli:2020xnv}
R.~B.~Ferroli, A.~Mangoni and S.~Pacetti,
Eur. Phys. J. C \textbf{80}, 903 (2020).

\bibitem{Wei:2009zzh}
D.~H.~Wei,
J. Phys. G \textbf{36}, 115006 (2009).

\bibitem{Jiao:2016syk}
J.~Jiao [BESIII],
PoS \textbf{CHARM2016}, 046 (2016).

\bibitem{Kivel:2022fzk}
N.~Kivel,
Eur. Phys. J. A \textbf{58}, 138 (2022).

\bibitem{Mangoni:2022yqq}
A.~Mangoni,
arXiv:2202.08542.

\bibitem{BaldiniFerroli:2019abd}
R.~Baldini Ferroli, A.~Mangoni, S.~Pacetti and K.~Zhu,
Phys. Lett. B \textbf{799}, 135041 (2019);

\bibitem{Micu:1968mk}
L.~Micu,
Nucl. Phys. B \textbf{10}, (1969).

\bibitem{LeYaouanc:1972vsx}
A.~Le Yaouanc, L.~Oliver, O.~Pene and J.~C.~Raynal,
Phys. Rev. D \textbf{8}, (1973).

\bibitem{Ackleh:1996yt}
E.~S.~Ackleh, T.~Barnes and E.~S.~Swanson,
Phys. Rev. D \textbf{54}, 6811-6829 (1996).

\bibitem{Simonov:2011cm}
Y.~A.~Simonov,
Phys. Rev. D \textbf{84}, 065013 (2011).

\bibitem{Weber:1988bt}
H.~J.~Weber,
Phys. Lett. B \textbf{218}, 267-271 (1989).

\bibitem{Chen:2007xf}
C. Chen, X. L. Chen, X. Liu, W. Z. Deng and S. L. Zhu,
Phys. Rev. D \textbf{75}, 094017 (2007).

\bibitem{Ke:2011wd}
H.~W.~Ke, Y.~Z.~Chen and X.~Q.~Li,
Chin. Phys. Lett. \textbf{28}, 071301 (2011).

\bibitem{Wang:2013lpa}
T.~Wang, G.~L.~Wang, H.~F.~Fu and W.~L.~Ju,
JHEP \textbf{07}, 120 (2013).

\bibitem{Gong:2021jkb}
K.~Gong, H.~Y.~Jing and A.~Zhang,
Eur. Phys. J. C \textbf{81}, 467 (2021).

\bibitem{Garcia-Tecocoatzi:2022zrf}
H.~Garcia-Tecocoatzi, A.~Giachino, J.~Li, A.~Ramirez-Morales and E.~Santopinto,
arXiv:2205.07049.

\bibitem{ParticleDataGroup:2022pth}
R.~L.~Workman \textit{et al.} [Particle Data Group],
PTEP \textbf{2022}, 083C01 (2022).

\bibitem{Xia:2021agf}
L.~Xia, C.~Rosner, Y.~D.~Wang, X.~Zhou, F.~E.~Maas, R.~B.~Ferroli, H.~Hu and G.~Huang,
Symmetry \textbf{14}, no.2, 231 (2022).

\bibitem{Liu:2022pdk}
C.~W.~Liu and C.~Q.~Geng,
arXiv:2205.08158.

\bibitem{Geng:2020ofy}
C.~Q.~Geng, C.~W.~Liu and T.~H.~Tsai,
Phys. Rev. D \textbf{102}, 034033 (2020);
J.~Zhang, X.~N.~Jin, C.~W.~Liu and C.~Q.~Geng,
Phys. Rev. D \textbf{107}, 033004 (2023);
C.~Q.~Geng, X.~N.~Jin, C.~W.~Liu, X.~Yu and A.~W.~Zhou,
Phys. Lett. B \textbf{839}, 137831 (2023).

\bibitem{Jin:2021onb}
X.~N.~Jin, C.~W.~Liu and C.~Q.~Geng,
Phys. Rev. D \textbf{105}, no.5, 053005 (2022).

\bibitem{Segovia:2012cd}
J.~Segovia, D.~R.~Entem and F.~Fern\'andez,
Phys. Lett. B \textbf{715}, 322-327 (2012).

\bibitem{BaBar:2011btv}
J.~P.~Lees \textit{et al.} [BaBar],
Phys. Rev. D \textbf{86}, 012008 (2012);
M.~Ablikim, J.~Z.~Bai, Y.~Bai, Y.~Ban, X.~Cai, H.~F.~Chen, H.~S.~Chen, H.~X.~Chen, J.~C.~Chen and J.~Chen, \textit{et al.}
Phys. Lett. B \textbf{693}, 88-94 (2010).

\end{thebibliography}
\end{document}